\documentstyle[prl,aps,epsfig]{revtex}

\begin{document}

\draft

\title{Scale-free networks are not robust under neutral evolution}

\twocolumn[\hsize\textwidth\columnwidth\hsize\csname@twocolumnfalse\endcsname

\author{Michael H\"ornquist\cite{email}
	\cite{address}}

\address{NORDITA \\
	Blegdamsvej 17, DK-2100 Copenhagen, Denmark}

\date{\today}
     
\maketitle

\begin{abstract}
Recently it has been shown that a large variety of different
networks have power-law (scale-free) distributions
of connectivities.
We investigate the robustness of such a distribution in discrete 
threshold networks under neutral evolution. 
The guiding principle for this is robustness in the resulting
phenotype.
The numerical results show that a power-law distribution is
not stable under such an evolution, and the network approaches
a homogeneous form where the overall distribution of
connectivities is given by a Poisson distribution.
\end{abstract}

\pacs{87.10.+e, 05.65.+b, 87.23.-n}

\vskip2pc]
\narrowtext      

There are many different areas of contemporary science where the
concept of networks are of special importance.
Recently, the amazing result that such diverse networks as
the World Wide Web \cite{WWW}, collaborations of movie actors 
\cite{Barabasi-99}, the electrical
power grid of western USA \cite{Barabasi-99}, citation patterns of scientific 
publications \cite{Redner-98} and metabolic networks
\cite{Jeong-00}, all
have distributions of  connectivities that are scale-free,
i.e., obey some power-law.
One reason for this quite general behaviour seems to be that
the networks are grown by addition of new nodes
and that each such new node preferentially attaches to other
nodes with high number of connections \cite{Barabasi-99}.
These networks are also more robust against unintelligent attacks than
homogeneous networks, where each node has approximately the
same number of connections \cite{Albert-00}.
For the case of genetic regulatory networks,
mutation is an example of such an ``unintelligent
attack'', and it is perhaps not surprising that the same type
of scale-free distributions are found in the metabolic networks
of so far 43 different organisms
from all three domains of life (bacteria, eukarya, and archaea)
\cite{Jeong-00}.

The evolution of life is a random process with selection
\cite{Darwin}, although all details about how this occur
with interactions among genotypes, phenotypes, and
environment are not totally clear.
Neutral evolution is the hypothesis that evolution
mainly proceeds as a random walk which does not affect the
phenotypes \cite{Kimura}.
Experimentally, it is supported on the microlevel by the fact that
most of the important macromolecules of life have forms
which are functionally identical variants.
If this idea of neutrality is correct also on a higher level, it means that
the usefulness of fitness landscapes for describing the evolution
of life as a hill climbing process is limited.

Here we explore the idea of neutral evolution and how the
distribution of connectivities in a genetic regulatory 
network changes under
mutations that are phenotypically silent.
The fundamental constituents in our model are the genes of the organism,
represented by the nodes of the network.
It has been suggested that such a system can be well approximated
by a Boolean network \cite{Kauffman}, because of the ``on-off'' nature
of the biochemical switches.
The exploration is performed by simulating evolution in discrete threshold
networks with robustness as the guiding principle for when
a mutation will survive to future generations.
We find by numerical simulations that the scale-free
distribution cannot be maintained under neutral evolution
in such networks. Instead, the networks evolve towards a
Poisson distribution, regardless of the actual realization of
the  initial scale-free distribution.


The discrete threshold network is composed of $N$ nodes,
$\sigma_i$, which are connected by a square matrix with elements
$w_{ij}$.
The values of the nodes are $\sigma_i\in \{-1,1\}$,
representing the corresponding gene to be expressed $(+1)$ or 
not $(-1)$.
The coupling matrix takes values $w_{ij} \in \{-1,0,1\}$,
with $+1$ if gene $j$ is an activator of gene $i$, $-1$ if
it is a repressor, and $0$ if no connection exist.
The dynamics of the network is described by the updating
rule
\begin{equation}
\sigma_i(t+1) = \text{sgn}\left(\sum_{j=1}^N w_{ij} \sigma_j(t)\right),
\end{equation}
where the sign-function is defined as $-1$ for all negative arguments,
and $+1$ otherwise (including zero).
Of special importance is the mean number of connections to each node,
$\bar{K}$, which is calculated as
\begin{equation}
\bar{K}=\frac{1}{N} \sum_{i=1}^N \sum_{j=1}^N |w_{ij}|,
\label{Kmean}
\end{equation}
that is, we make no distinction between repressors and activators,
and it has the same value for both ingoing and outgoing connections.

This is a special case of Boolean networks, with similar
structural and statistical properties \cite{Kurten-88}.
These properties include both transients and limit cycles (attractors),
as well as phase transitions for a specific critical connectivity, 
$\bar{K}=K_c$.
An analytical approach is limited due to the
non-Hamiltonian character of the system, but results within
the so-called \emph{annealed approximation} show, for Boolean networks,
that for $\bar{K}$ below the critical connectivity $K_c = 2$, 
there are many
disconnected regions, while above $K_c$ most of the nodes are
connected and the limit cycle period increases exponentially
with the number of nodes.
Also, at least in some intervals above $K_c$, the size of the
attractors diverges almost exponentially with increasing connectivity
\cite{Derrida-86}.
Note here that we use a form where the number of ingoing connections might
differ from the number of outgoing, i.e., the matrix does not have
to be symmetric, and that we do not impose the restriction that the number
of connections should be the same for all nodes \cite{Kurten-88}.

If every two nodes in a network are connected with the same
probability $p$, we have the Erd\H{o}s-R\'enyi model 
for random graphs \cite{ER}, sometimes referred to as a
\emph{homogeneous} network.
It is well known that in such networks, the number of
connections to each node follows a Poisson distribution
(with exponential decaying tails), 
and hence there will hardly be any
nodes with a large number of connections.
Loosely speaking, this is the most common form of a
random graph.

\begin{figure}
\centering \epsfig{figure=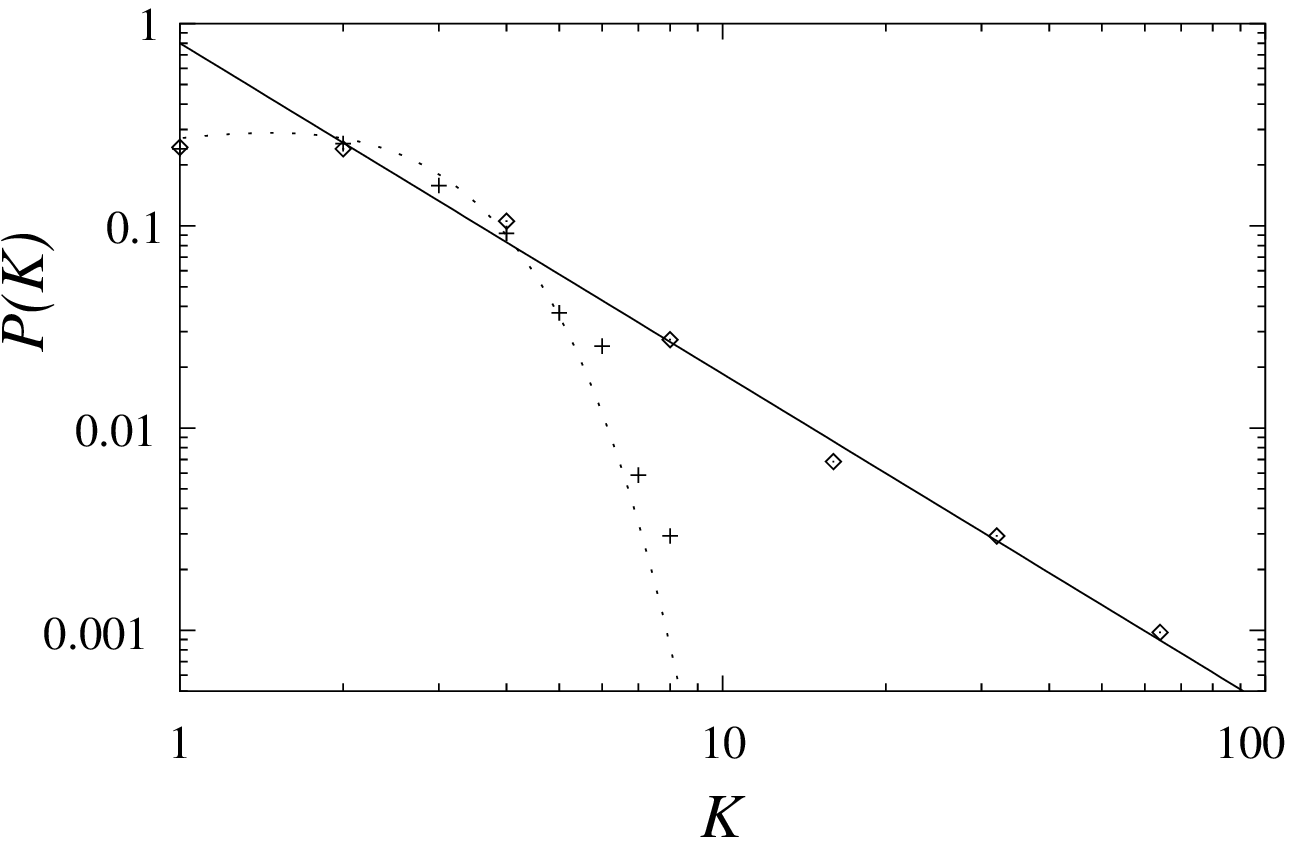,width=0.95\linewidth}
\caption{Distribution functions for number of ingoing connections
	for an initially scale-free
	network with $N=1024$ nodes at two different generations.
	Diamonds: Initial power-law distribution.
	Crosses: Distribution after 30 000 generations of
	neutral evolution.
	The full line is a power-law, $P(K)\sim K^{-\gamma}$ with
	$\gamma=1.63$,  and the dotted line is
	a Poisson distribution (see text for details).} 
\label{fig:variation}
\end{figure}

To simulate neutral evolution, we start by generating a network with elements
$w_{ij}$ by the procedure described in \cite{Barabasi-99}, i.e.,
we start from a small random network, and add new nodes by preferential
attachment.
The result is a scale-free network 
with a probability for a given node to have $K$ links 
(either ingoing or outgoing) proportional to 
$K^{-\gamma}$ for $K\ge 2$, with, in our case, $\gamma\approx 1.63$ 
(solid line in Fig.\ \ref{fig:variation}).
We use consequently in this paper $N=1024$, which results in
an average connectivity value initially slightly below 
the critical value $K_c=2$.
Interestingly, a recent letter showed that another form
for evolving a discrete threshold network (adding links to
quiet nodes, removing links from active) leads to an
average connectivity of 2.55 for this size of network \cite{Bornholdt-00}.
The sign of a specific connection specifies if we have
an activator (positive value) or a repressor (negative value).
These signs are here chosen randomly with equal probabilities.

The evolution now proceeds by the following procedure:
The network is mutated by either 
\begin{enumerate}
\item One non-zero element is put to zero (connection removed)
\item One zero element is turned into $\pm 1$ (with equal probability)
	(connection added)
\item Both of the above, i.e., one connection is added and another removed
\end{enumerate}
These three alternatives (in the order given) occur with the probabilities
$0.300:0.333:0.367$, forming a new, mutated network.
The values of these probabilities were chosen to obtain a network
with a relatively constant mean number of connections also for
the comparatively small number of generations and the initially
low number of mean connectivity we consider.
However, also other values have been tried, and the results do
not depend critically on their exact magnitudes.
To either reject or accept the new, mutated network, 
we use robustness as the guiding principle.
This is achieved by
picking by random an initial state, $\{\sigma_i\}$, with equal probability 
for each single node $\sigma_i$ being either positive or negative.
This state is iterated in both the original and the mutated
network until we either enter into the same limit cycle in both
networks, or the two iterants cease to coincide.
In the former case we accept the mutated network, and replace the
original one with the mutated version. 
This is then an evolutionary step within the neutral evolution.
In the latter case, we reject the new, mutated network, since the
effect of the mutation were not silent.
Finally, we return to the mutation step and repeat the procedure.
Notice this introduces two different time scales in the evolution.
The one corresponding to the iteration of states $\{\sigma_i\}$
relates to a single generation, while the much slower
process of accepting or rejecting new networks, i.e.,
the rewiring of connections $w_{ij}$, corresponds to
the evolution over generations.

This way of simulating neutral evolution has earlier been explored
by Bornholdt and Sneppen, both by truly Boolean networks
\cite{Bornholdt-98} and by discrete threshold networks 
\cite{BornholdtSneppen-00}.
They studied, however, the phenomenon of punctuated equilibrium and
distribution of waiting times, and ignored the distribution 
of connections.
Their study clarified that this model exhibit
many of the known properties of evolution, such as
$1/f$ power spectra \cite{Sole-97} and $1/t^2$ stability distribution
\cite{Sneppen-95}, in 
accordance with similar scalings found in the statistics
of birth and death in the evolutionary record.


In Fig.\ \ref{fig:variation} we show the initial 
distributions of connections
and the distribution after 30 000 generations
in one evolutionary run
for the number of connections leading \emph{into} the nodes.
The result for connections leading \emph{out} from them
are quite similar, and are for clarity not shown.
Although the limited number of nodes
(due to computational constraints) makes the statistics
somewhat fuzzy, it is still clear that the distribution changes
from a power-law to an approximately Poisson distribution.
To get a better picture, we have used the well-established
technique of binning the values for the initial distribution.
The solid and dotted lines  are a power-law and a
Poisson distribution, respectively.
The exponent of the straight line $P(K) \sim K^{-\gamma}$ is 
found by a least squares fit to be  $\gamma \approx 1.63$.
This should be compared to the theoretical value given in
\cite{Barabasi-99} for a fully directed graph of infinite size, which is 
$\gamma=2$.
Although the number of nodes we use is small compared to
the networks considered there, the correspondence seems
acceptable.
The Poisson distribution drawn is for the expectation value estimated
by the mean value of connectivities, $\bar{K}$, at the actual generation.
This means that in the general Poisson distribution function
\begin{equation}
P(K)=\frac{\mu^K}{K!} \exp(-\mu),
\label{Poisson}
\end{equation}
we estimate the parameter $\mu$ with the mean connectivity $\bar{K}$.
No curve fitting is used this time, but nevertheless the correspondence
is quite striking.

\begin{figure}
\centering \epsfig{figure=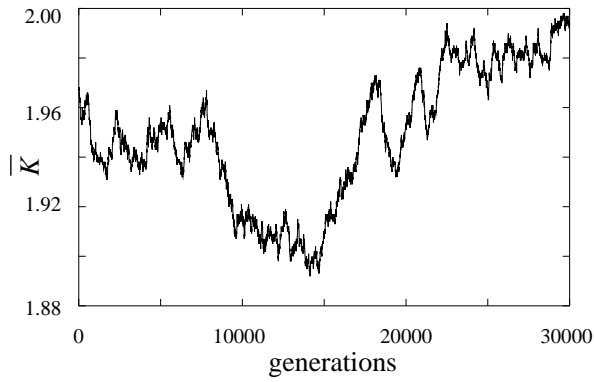,width=0.95\linewidth}
\caption{Mean connectivity for an initially scale-free network evolving
	under neutral evolution.}
\label{fig:connectivity}
\end{figure}

\begin{figure}
\centering \epsfig{figure=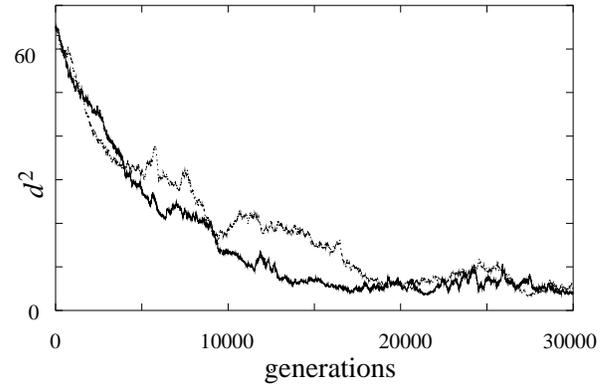,width=0.95\linewidth}
\caption{Weighted mean square deviations from the Poisson distribution
	for an initially scale-free network, evolving
	under neutral evolution, where the expectation value
		at each generation is estimated by the actual mean
		connectivity. Full line represents ingoing connections,
		dotted line represents outgoing connections.}
\label{fig:Poisson}
\end{figure}

In Fig.\ \ref{fig:connectivity}, we show the variation of mean connectivity,
$\bar{K}$, for the first 30 000 generations. 
Because of the definition (\ref{Kmean}), this is the same value
both for outgoing and ingoing connections.
The curve shows  that
the mean number of connections remains fairly constant,
although the detailed dynamics is non-trivial
(with punctuated equilibria, etc., as discussed in
\cite{Bornholdt-98,BornholdtSneppen-00}).
It also shows that we in this run constantly are below the critical value
of $K_c=2$.
This is, however, not a critical aspect of the simulation, which other
runs (not shown) have indicated.
Hence any change in distribution among these connections cannot be due to
changes in the mean connectivity.
Nevertheless, there is according to Fig.\ \ref{fig:variation} a real
change in distribution from the start of the simulated evolution to the end
of our calculations.
To shed some further light onto the transition from a scale-free
network with a power-law distribution to a homogeneous network with
a Poisson distribution, we calculate the weighted mean square deviation
\begin{equation}
d^2 = \frac{1}{N} \sum_{K=1}^{N} K |n(K) - N P(K)|^2,
\label{ms-weight}
\end{equation}
where $n(K)$ is the number of nodes with $K$ connections and
$P(K)$ is the Poisson distribution (\ref{Poisson}).
For each comparison, we use the actual value of $\bar{K}$ as 
estimate for the expectation value $\mu$, but no other fitting is performed.
Because the tail of the distribution function is the most critical, 
we give higher weight to larger number of connections
by multiplying each term with the actual number of connections.
The results are shown in
Fig.\ \ref{fig:Poisson}.
Although the exact details for the ingoing  and outgoing connections
differ,
and it is clearly seen that the distributions eventually approach the
Poisson distribution, regardless of which measure we consider.


The lengths of the limit cycles for the accepted
networks in this evolutionary scenario
vary between 1 and 20, with an average of approximately 6.
The transients have lengths between 10 and 35 steps.
Both these results indicate that we are in the regime of many, small,
disconnected attractors, which is fully consistent with the
mean connectivity $\bar{K}$ being less than the critical
value $K_c=2$.

\begin{figure}
\centering \epsfig{figure=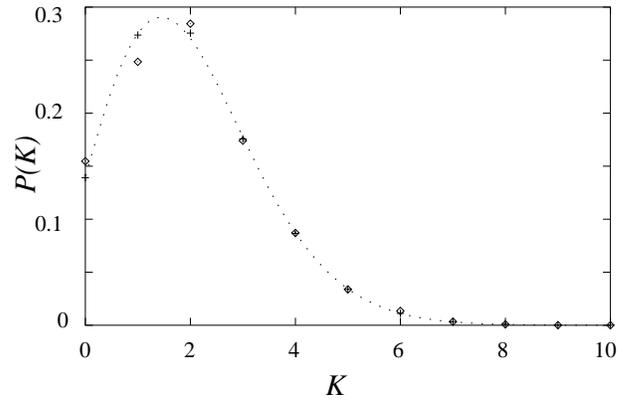,width=0.95\linewidth}
\caption{Mean distribution for all generations between 200 000 and 300 000 for
	an initially scale-free network evolving
	under neutral evolution with constant connectivity.
	Diamonds are ingoing connections and crosses outgoing.
	The dotted line is a Poisson distribution with expectation
	value estimated as the mean connectivity.}
\label{fig:konst}
\end{figure}

Better statistics, i.e., less fuzzy distributions, 
are obtained if we change the mutation 
rule above somewhat.
Instead of having the possibility to separately add or remove
a connection, we stick solely to alternative three, which
means that the \emph{number} of connections, and hence $\bar{K}$,
remains constant.
This is clearly a less  realistic scenario than
before from a biological point of view, 
but can help to see what distributions we really have.
In Fig.\ \ref{fig:konst}, we show the mean values of the distributions
for all generations between number 200 000 and 300 000,
when we start with the same initial scale-free network used to obtain
the results of Fig.\ \ref{fig:variation}.
The dotted line is the theoretical Poisson distribution for the
actual mean connectivity $(\bar{K}=1.9678)$.
The weighted mean square deviations from this theoretical value 
are $1.18$ for the ingoing connections and $0.06$ 
for the outgoing, respectively.
To the prize of having incorporated an unrealistic restriction,
we have obtained distributions which are considerably
closer to the theory.

To check the robustness of these results, 
we have repeated the calculations many times with different forms
of initial network in the construction of the original scale-free
network, as well as checked many different realizations.
We have also started with
networks with a power-law distributed number of connections obtained
directly from a random number generator,
i.e., without the process described in \cite{Barabasi-99}.
In all these cases, our results do not change, i.e., the systems
always end up with a Poisson distribution consistent with
the random graph theory.


In conclusion, we have studied the evolution of initially
scale-free networks, i.e., networks where the distribution of
the number of connections to each node follows a power-law.
The networks are evolved under the hypothesis of neutral evolution, 
which is implemented as a robustness criterium for the limit cycles
in discrete threshold networks.
The result is that the scale-free distribution is not robust
under such an evolution, but instead all networks end up
in a homogeneous form, with a Poisson distribution of
connectivities.
This result  is surprising, since it has been shown that
a scale-free network is more robust towards random attacks than
an exponential network \cite{Albert-00}.
Although one should be very careful with infering any definite
statements with respect to biology from such simple models as
the one presented here, we can speculate and draw the tentative conclusion
that the addition of new nodes with preferential attachment
seems to be a force that manages to repress the changes due
to neutral evolution.
Future studies might shed some light on the
presumably different time-scales that are active here.

The author would like to thank Kim Sneppen for useful discussions
on discrete threshold networks, 
Ingve Simonsen for careful reading of the manuscript,
and Mattias Wahde for pointing out the results on scale-free
networks.
Financial support from The Swedish Foundation for International Cooperation in
Research and Higher Education , STINT, is also gratefully
acknowledged.

\clearpage

\newpage

\end{document}